\documentclass[superscriptaddress,twocolumn,showpacs,preprintnumbers,amsmath,amssymb]{revtex4}

\usepackage [dvips]{epsfig}
\usepackage {graphics}
\usepackage {revsymb}
\usepackage{graphicx}
\usepackage{dcolumn}
\usepackage{bm}

\begin{document}


\title{First-principles study of the electrooptic effect in
ferroelectric oxides}

\author{Marek Veithen}
\affiliation{D\'epartement de Physique, Universit\'e de Li\`ege,
B-5, B-4000 Sart-Tilman, Belgium}

\author{Xavier Gonze}
\affiliation{Unit\'e PCPM, UCL,
            place Croix du Sud, 1, B-1348 Louvain-la-Neuve, Belgium}

\author{Philippe Ghosez}
\affiliation{D\'epartement de Physique, Universit\'e de Li\`ege,
B-5, B-4000 Sart-Tilman, Belgium}

\date{\today}

\begin{abstract}

We first present a method to compute 
the electrooptic tensor from
first principles, explicitly taking into account the electronic, ionic and
piezoelectric contributions. We then study the non-linear optic behavior
of three paradigmatic ferroelectric oxides.
Our calculations reveal the dominant contribution of the
soft mode to the electrooptic coefficients in LiNbO$_3$ and BaTiO$_3$
and its minor role in PbTiO$_3$. We identify the coupling between the
electric field and the polar atomic displacements along the B-O chains
as the origin of the large electrooptic response in perovskite 
ABO$_3$ compounds.

\end{abstract}



\maketitle


The electrooptic (EO) effect describes the change of refractive index
of a material in a static electric field and is exploited in
various technological applications~\cite{lines}. Ferroelectric
ABO$_{3}$ compounds exhibit unusually large EO coefficients and are
therefore materials of choice for optical devices. 
Since the
seventies, LiNbO$_3$ EO modulators have been widely used in fiber-optic
transmission systems~\cite{ieee}. More recently, there has been increasing
interest in epitaxially grown BaTiO$_3$ thin films for optical
waveguide modulators~\cite{apl81_1375}. The EO effect is the origin of
the photorefractive effect, exploited in non-volatile holographic data
storage in LiNbO$_3$~\cite{photore1}. It was also used to probe locally
the ferroelectricity in SrTiO$_3$ films~\cite{prl89_147601}.

Finding better EO materials is a desirable goal. However, the
experimental characterization of optical non-linearities requires
high-quality single crystals that are not always directly accessible
nor easy to make. Input from accurate theoretical calculations allowing
to {\it predict} the non-linear optical behavior of crystalline solids would
therefore be particularly useful.

For many years, theoretical investigations of non-linear optical phenomena were
restricted to semi-empirical approaches such as shell
models~\cite{ferroel234_61} or bond-charge
models~\cite{prb7_2600,jpc15_825}.
In the last decade, significant theoretical advances have been reported
concerning first-principles density functional theory (DFT)
calculations of the behavior of periodic systems in an external electric
field~\cite{prb63_155107,prl89_117602}
and opened the way to direct predictions of various
optical phenomena. Recently, particular attention has been paid
to the calculation of non-linear optical (NLO) susceptibilities
and Raman cross sections~\cite{prb53_15638_prb66_100301,prl90_36401}.

In this Letter, we go one step further and present a method to predict
the {\it linear EO coefficients} of periodic solids within DFT.
Our method is very general, and can be applied to paradigmatic
ferroelectric oxides : LiNbO$_3$, BaTiO$_{3}$ and PbTiO$_{3}$.
We find that first-principles calculations are fully predictive, and provide
significant new insights into the microscopic origin of the EO effect.
In particular, we highlight the {\it predominent role of the soft mode} in
the EO coupling of LiNbO$_3$ and BaTiO$_{3}$, in contrast with
{\it its minor role} in
PbTiO$_{3}$.


At linear order, the dependence of the optical dielectric tensor
$\varepsilon_{ij}$
on the static (or low-frequency) electric
field $\mathcal{E}_{\gamma}$ is described by the linear EO tensor
r$_{ij \gamma}$:
\begin{equation} \label{eq_eo_def}
\Delta \left ( \varepsilon^{-1} \right )_{ij}
=
\sum_{\gamma = 1}^{3}
r_{ij \gamma} \mathcal{E}_{\gamma}.
\end{equation}
Throughout this paper, we follow the convention
of using Greek and Roman indexes (resp.) to label static and optical fields
(resp.). We write all vector and tensor components in the
system of cartesian coodinates defined by the principal axes of the crystal
under zero field. We also refer to the atomic displacements
$\tau_{\kappa \alpha}$ [$\kappa$ labels an atom
and $\alpha$ a cartesian direction] within
the basis defined by the zone-center transverse optic (TO) phonon
eigendisplacements $u_m(\kappa \alpha)$:
$
\tau_{\kappa \alpha} =
\sum_m
\tau_m u_m(\kappa \alpha)
$.


Let us first consider the {\it clamped} (zero strain) EO tensor, $r_{ij
\gamma}^{\eta}$, in which all electric-field induced macroscopic
strains $\eta$ are forbidden.
This is achieved experimentally by working at a
frequency sufficiently high to avoid strain relaxations but
low compared to the frequency of the TO modes. Within the
Born-Oppenheimer approximation, we express the {\it total} derivative
of $\varepsilon_{ij}$ as the sum of two {\it partial} derivatives with
respect to $\mathcal{E}_{\gamma}$ and $\tau_m$:
\begin{equation} \label{eq_totder}
\frac{d \varepsilon_{ij}}{d \mathcal{E}_{\gamma}}
=
\left .
\frac{\partial \varepsilon_{ij}}{\partial \mathcal{E}_{\gamma}}
\right |_{\tau=0}
+
4 \pi \sum_{m}
\left .
\frac{\partial \chi_{ij}^{(1)}}{\partial \tau_{m}}
\right |_{\mathcal{E}=0}
\frac{\partial \tau_{m}}{\partial \mathcal{E}_{\gamma}}.
\end{equation}
The derivative in the first term of the righthand side of Eq.
(\ref{eq_totder}) is computed at clamped atomic positions.
It describes the {\it electronic} contribution to the EO tensor
and is proportional to the NLO susceptibilities
$\chi_{ijl}^{(2)}$.
The second term represents the {\it ionic} contribution.
It depends on the first-order change of the linear dielectric
susceptibility due to atomic displacements, and is related to the Raman
susceptibility
$
\alpha_{ij}^m =
\sqrt{\Omega} \sum_{\kappa \alpha}
\frac{\partial \chi_{ij}^{(1)}}{\partial \tau_{\kappa \alpha}}
u_m(\kappa \alpha)
$
of mode $m$ [$\Omega$ is the unit cell volume],
as well as to the amplitude of the ionic relaxation
induced by the field $\mathcal{E}_{\gamma}$.
$\partial \tau_m / \partial
\mathcal{E}_{\gamma}$ can be expressed in terms of (i) the TO phonon
mode frequencies $\omega_m$ and (ii) the TO mode polarities
$
p_{m,\gamma} =
\sum_{\kappa,\beta} Z^{\ast}_{\kappa, \gamma \beta}
u_m(\kappa \beta),
$
directly linked to the infrared (IR) intensities
\cite{xgcl}.
Combining this
with the previous equations, we obtain
the clamped EO tensor~\cite{unpublished}
\begin{equation} \label{eq_ijgamma}
r_{ij \gamma}^{\eta}
=
\left .
\frac{-8 \pi}{n_i^2 n_j^2}
\chi_{ijl}^{(2)}
\right |_{l = \gamma}
-
\frac{4 \pi}{n_i^2 n_j^2 \sqrt{\Omega}}
\sum_m
\frac{\alpha_{ij}^m p_{m,\gamma}}{\omega_m^{2}}
\end{equation}
where $n_i$ and $n_j$ are the principal
refractive indices.


Let us now consider the {\it unclamped} (zero stress) EO tensor,
$r_{ij \gamma}^{\sigma}$. It can be shown ~\cite{unpublished} that
the macroscopic
expression proposed in Ref. \cite{prb50_5941} is still valid at
the microscopic level so that the the piezoelectric contribution
to $r_{ij \gamma}^{\sigma}$ can be computed from the elasto-optic
coefficients p$_{ij \alpha \beta}$ and the piezoelectric strain coefficients
d$_{\gamma \alpha \beta}$
\begin{equation} \label{eq_unclamped}
r_{ij \gamma}^{\sigma} = r_{ij \gamma}^{\eta}
+ \sum_{\alpha, \beta = 1}^{3} p_{ij \alpha \beta} d_{\gamma \alpha \beta}.
\end{equation}

An expression similar to Eq. (\ref{eq_ijgamma}) was previously
used by Johnston~\cite{prb1_3494} to estimate the clamped EO
tensor of LiNbO$_3$ and LiTaO$_3$ from IR
and Raman measurements. However, this semi-empirical approach was limited by the
indeterminacy of the relative sign of $\bm{p}_m$ and
$\bm{\alpha}^m$. As discussed below, the direct evaluation
of Eq. (\ref{eq_ijgamma}) and (\ref{eq_unclamped}) from first
principles provides an {\it easier} and {\it more accurate} estimate
of the EO tensor.


We have implemented this formalism in the {\sc abinit}
open software~\cite{abinit}, within
the local density approximation (LDA) to the DFT.
The optical
dielectric tensor, Born effective charges, phonon frequencies and
eigendisplacements are computed from
linear response~\cite{xgcl}.
The piezoelectric strain coefficients d$_{\gamma \alpha \beta}$
are deduced from the piezoelectric stress coefficients
e$_{\gamma \alpha \beta}$ and the elastic constants.
These two quantities, as well as the elasto-optic tensor p$_{ij \alpha \beta}$,
are obtained from finite differences.
The non-linear response functions $\chi^{(2)}_{ijl}$
and $\partial \chi^{(1)}_{ij}/\partial \tau_{\kappa \alpha}$
are computed from a perturbative approach using a new
implementation based on the $2n + 1$ theorem.
To reach reasonable k-point sampling convergence,
we combined the recently proposed PEAD expression~\cite{prb63_155107}
and the finite difference
formula of Marzari and Vanderbilt~\cite{prb56_12847}
to compute the perturbation expansion of the polarization.
More details 
will be provided elsewhere~\cite{unpublished}.
The method was tested on various cubic semiconductors and
provides results in close agreement with earlier
studies~\cite{prl89_117602,prb53_15638_prb66_100301}.

For BaTiO$_3$ and PbTiO$_3$, we use extended norm-conserving
pseudopotentials~\cite{prb48_5031}, a planewave
kinetic energy cutoff of 45 hartree and a $10 \times 10 \times 10$
k-point grid.
For LiNbO$_3$, we use the same norm-conserving
pseudopotentials as in Ref. \cite{prb65_214302} as well as the Born effective
charges, phonon frequencies and eigenvectors already reported in that
paper.  For this compound, a $8 \times 8 \times 8$ k-point grid
and a planewave kinetic energy cutoff of 35 hartree give converged values for
$\chi^{(2)}_{ijl}$ and
$\partial \chi^{(1)}_{ij}/\partial \tau_{\kappa \alpha}$.



First, we study LiNbO$_3$.
This compound has a trigonal symmetry with 10 atoms
per unit cell. The theoretical lattice constants and atomic
positions are reported in Ref. \cite{prb65_214302}.
LiNbO$_3$ undergoes a single transition at 1480 K
from a centrosymmetric high-T paraelectric $R\overline{3}c$ phase
to a ferroelectric low-T R3c
ground state.  The form of the EO tensor
depends on the choice of the cartesian axes.
Here, we follow the I.R.E. Piezoelectric
Standards~\cite{jpc2_855}.


With this choice of axes, the EO tensor in the ferroelectric phase of
LiNbO$_3$
has 4 independent elements (Voigt notations):
r$_{13}$, r$_{33}$, r$_{22}$ and r$_{51}$.
The TO modes
can be classified into
$4 A_1 + 5 A_2 + 9 E$. The A$_1$ and E modes are 
Raman and IR
active. Only the A$_1$ modes couple to
r$_{13}$ and r$_{33}$, while the E modes are linked to r$_{22}$ and r$_{51}$.
Table \ref{tab_linbo3} gives these four clamped
coefficients~\cite{note_rc}, as well as the contribution of
each optical phonon.
For comparison, we mention the coefficients computed by
Johnston~\cite{prb1_3494} from measurements of IR and Raman
intensities (IR + R) as well as the results of a bond-charge model (BCM)
calculation by Shih and Yariv~\cite{jpc15_825}.
The first-principles calculations correctly predict
the sign of the four EO coefficients~\cite{jpc2_855}.
The absolute values are also well reproduced by our method,
especially if we take into account the fact that NLO properties
are generally difficult to determine accurately. The experimental values
are sensitive to external parameters such as temperature
changes \cite{jap65_2406} and the stoichiometry of the samples. For example,
using crystals of various compositions,
Abdi and co-workers measured absolute values between 1.5 pm/V
and 9.9 pm/V for r$_{22}^{\sigma}$ \cite{jap84_2251}.
These difficulties support the need for sophisticated theoretical
tools to predict NLO properties.
In contrast to the models of Refs. \cite{prb1_3494,jpc15_825},
our method is predictive and does not use any experimental parameters.
Moreover, it reproduces r$_{13}^{\eta}$, r$_{33}^{\eta}$
and r$_{22}^{\eta}$ better than the semiempirical models.

Our approach also provides some insight into the origin of the high
LiNbO$_{3}$ EO
response.
All EO coefficients are dominated by the ionic contribution of
the A$_1$ TO1
and the E TO1 modes.
This can be explained as follows.
At the paraelectric-ferroelectric phase transition, the
unstable A$_{2u}$ and E$_u$ modes of the paraelectric phase
transform to {\it low-frequency} and
{\it highly polar} modes in the ferroelectric phase~\cite{prb65_214302},
generating a large EO response if they exhibit, in addition, a
{\it large Raman susceptibility}.
The A$_1$ TO1 and E TO1 modes of the ferroelectric phase
have a strong overlap of respectively 0.82 and 0.68
with the unstable A$_{2u}$ and E$_u$ modes of the paraelectric phase
and combine giant polarity~\cite{prb65_214302}
and large Raman susceptibility (see below
for the A$_1$ mode).

In Table \ref{tab_linbo3}, we also report the
unclamped EO coefficients in LiNbO$_3$.
As the piezoelectric coefficients d$_{31}$ (-1 p$C/N$) and
d$_{33}$ (6 p$C/N$) are small compared
to d$_{15}$ (55.9 p$C/N$)
and d$_{22}$ (21.6 p$C/N$),
the piezoelectric effect is important for
r$_{22}^{\sigma}$ and r$_{51}^{\sigma}$
and negligible for
r$_{13}^{\sigma}$ and r$_{33}^{\sigma}$.
The unclamped EO coefficient r$_{51}^{\sigma}$
is nearly twice as large as the clamped one. Moreover, its theoretical
value is in better agreement with
the experiment than that of the clamped one.
This suggests that the piezoelectric contribution was not entirely
eliminated during the measurement of r$_{51}^{\eta}$; the
correct value of the clamped coefficient might be closer to the
theoretical 14.9 pm/V.

\begin{table}[tbp]
\caption{\label{tab_linbo3}
EO tensor (pm/V) in LiNbO$_3$ :
electronic, ionic and piezoelectric
contributions, and comparison with experiment,
for the clamped and unclamped cases. The ionic part is
split into contributions from
TO modes ($\omega_m$ in cm$^{-1}$).
}
\begin{ruledtabular}
\begin{tabular}{lr|rrrrr|rrr}
& & & \multicolumn{3}{c}{A$_1$-modes} & &
         \multicolumn{3}{c}{E-modes} \\
& & & $\omega_m$ & r$_{13}$ & r$_{33}$ & &
         $\omega_m$ & r$_{22}$ & r$_{51}$ \\
\hline
Electronic & & &         & 1.0 & 4.0 & &     & 0.2 & 1.0   \\
Ionic     &TO1   & & 243 & 6.2 &18.5 & & 155 & 3.0 & 7.5   \\
          &TO2   & & 287 &-0.2 &-0.4 & & 218 & 0.4 & 1.5   \\
          &TO3   & & 355 &-0.1 & 0.0 & & 264 & 0.6 & 1.3   \\
          &TO4   & & 617 & 2.8 & 4.8 & & 330 &-0.3 & 1.2   \\
          &TO5   & &     &     &     & & 372 &-0.2 & 0.4   \\
          &TO6   & &     &     &     & & 384 &-0.1 &-0.2   \\
          &TO7   & &     &     &     & & 428 & 0.2 & 0.2   \\
          &TO8   & &     &     &     & & 585 & 0.7 & 2.1   \\
          &TO9   & &     &     &     & & 677 & 0.0 & 0.0   \\
\multicolumn{2}{r|}{Sum of ionic}  
                 & &     & 8.7 &22.9 & &     & 4.4 &13.9   \\
Strain & & &             & 0.8 & 0.1 & &     & 3.0 &13.7   \\
\hline
Clamped &Present & &     & 9.7 &26.9 & &     & 4.6 &14.9   \\
             &Exp. \cite{rauber}
           & &    & 8.6  &30.8  & &     & 3.4  & 28     \\
             &IR+R \cite{prb1_3494}
           & &    & 12   & 39   & &     & 6    & 19     \\
             &BCM \cite{jpc15_825}
           & &    &      & 25.9 & &     &      & 20.5  \\
\hline
Unclamped  & Present & &  &10.5 & 27.0 & & & 7.5 & 28.6   \\
                & Exp. \cite{rauber}
           & & & 10.0  & 32.2  & & &  6.8  & 32.6    \\
                & Exp. \cite{jap84_2251}
           & & &       &       & & &  9.9 &         \\

\end{tabular}
\end{ruledtabular}
\end{table}


Second, we study PbTiO$_{3}$ and BaTiO$_{3}$.
Both compounds are stable at room temperature in a ferroelectric
distorted perovskite structure of tetragonal P4mm symmetry with
5 atoms per unit cell~\cite{Note-positions}.
In the P4mm phase, the TO modes can be classified
into $3 A_1 + 4 E + B_1$.
The EO tensor has only three independent elements: r$_{13}$,
and r$_{33}$, coupling to the A$_{1}$ modes, and r$_{42}$, linked
to the E modes. The B$_1$-mode is IR inactive
and does not influence the EO tensor.
The results are shown in Table~\ref{tab_eo}.

For PbTiO$_3$, we found
only measurements of r$_{13}^{\eta}$ and r$_{33}^{\eta}$,
which agree well with our theoretical results.
Moreover, our calculation predicts that
PbTiO$_{3}$ exhibits a large r$_{42}^{\eta}$,
in spite of its low r$_{33}^{\eta}$.
Combined with other
advantageous features, such as small thermo-optic
coefficients~\cite{jap77_2102},
this suggests that PbTiO$_{3}$ {\it might be} an interesting candidate for
EO applications {\it if properly oriented}.

In BaTiO$_{3}$, the low temperature structure is rhombohedral. The
P4mm phase is unstable and exhibits, in the harmonic
approximation, an unstable E-mode that prevents
the use of Eq. (\ref{eq_ijgamma}) to compute r$_{42}^{\eta}$.
The theoretical estimates of r$_{13}^{\eta}$ and r$_{33}^{\eta}$
are reasonably accurate 
despite
an underestimation of the theoretical r$_{33}^{\eta}$.
The origin of the error can be attributed to various sources.
First, the values computed for the P4mm phase correspond
to an extrapolation of the EO tensor to 0 K, while
experimental results are obtained at room temperature.
Also, linear and NLO susceptibilities can be
relatively inaccurate within the LDA. In this context,
note the use of the LDA
optical refractive indexes in Eq. (\ref{eq_ijgamma}),
overestimating the experimental values by about 10 \%.

\begin{table}[tbp]
\caption{\label{tab_eo} Electronic and ionic contributions of individual
TO modes ($\omega_m$ in cm$^{-1}$) to the clamped
EO tensor (pm/V) in the P4mm phase
of PbTiO$_3$ and BaTiO$_3$.
}
\begin{ruledtabular}
\begin{tabular}{lrrrrrrrrr}
      & \multicolumn{5}{c}{PbTiO$_3$} & & \multicolumn{3}{c}{BaTiO$_3$} \\
      & \multicolumn{3}{c}{A$_1$-modes} & \multicolumn{2}{c}{E-modes} & &
        \multicolumn{3}{c}{A$_1$-modes} \\
      &$\omega_m$ & r$_{13}^{\eta}$ & r$_{33}^{\eta}$ & $\omega_m$ &
r$_{42}^{\eta}$ & &
        $\omega_m$ & r$_{13}^{\eta}$ & r$_{33}^{\eta}$ \\
\hline
Elec. &      &  2.1 & 0.5 &      &  2.2 & &      & 1.0  & 2.1 \\
\hline
TO1   & 151  &  3.9 & 2.9 &  79  & 16.4 & & 161  & 1.0  & 1.0 \\
TO2   & 357  &  1.4 & 0.7 & 202  & 10.5 & & 300  & 5.7  &16.3 \\
TO3   & 653  &  1.6 & 1.8 & 269  &  0.2 & & 505  & 1.2  & 2.9 \\
TO4   &      &      &     & 484  &  1.2 & &      &      &     \\
\hline
Tot   &      &  9.0 & 5.9 &      & 30.5 & &      & 8.9  &22.3 \\
Exp. \cite{handbook_laser}
           &      & 13.8 & 5.9  \\
Exp. \cite{prb50_5941}
           &      &      &      &      &       & &      &10.2   & 40.6 \\
Exp. \cite{asss3_264}
           &      &      &      &      &       & &      & 8     & 28   \\
\end{tabular}
\end{ruledtabular}
\end{table}


We compare now the NLO response of the three compounds.
r$_{13}^{\eta}$ is similar for all of them,
while r$_{33}^{\eta}$ is significantly smaller in PbTiO$_{3}$
than in LiNbO$_3$ and BaTiO$_3$.
In the latter two compounds, the magnitude of r$_{33}^{\eta}$ is dominated
by one particular phonon mode. In BaTiO$_3$, the TO2 mode at 300 cm$^{-1}$
has a similar strong overlap (92\%) with the unstable mode in the
paraelectric phase than the TO1 modes in LiNbO$_3$, as previously
discussed. In PbTiO$_3$, all A$_1$ modes contribute almost equally
to r$_{33}^{\eta}$. The TO2 mode at 357 cm$^{-1}$ has the strongest
overlap (73\%) with the soft mode in the cubic phase. Surprisingly,
its contribution to r$_{33}^{\eta}$ is
{\it 23.5 times smaller} than the contribution of the 
TO2 mode in BaTiO$_3$.

To identify the origin of the distinctive behavior of PbTiO$_{3}$, we report
in Table \ref{tab_ramansus}
the mode polarities and
Raman susceptibilities of the A$_1$ TO modes. In the three
compounds, $\bm{\alpha}$ has two independent elements
$\alpha_{11}$ and $\alpha_{33}$
that determine the amplitude of r$_{13}^{\eta}$ and r$_{33}^{\eta}$.
$\alpha_{33}$ is large for the TO1 mode in LiNbO$_3$
and the TO2 mode in BaTiO$_3$. On the other hand, it is the smallest
for the TO2 mode in PbTiO$_3$, in agreement with experiments
\cite{jpc3_8695}.
Combined with a higher frequency
($\omega^2_{PbTiO_3}/\omega^2_{BaTiO_3} = 1.41$),
a lower polarity
($p_{BaTiO_3}/p_{PbTiO_3} = 1.49$),
and a larger value of the refractive index
($n^4_{PbTiO_3}/n^4_{BaTiO_3} = 1.35$),
this weak Raman susceptibility
($\alpha_{BaTiO_3}/\alpha_{PbTiO_3} = 8.27$)
explains the weak contribution of
the TO2 mode to r$_{33}^{\eta}$ in PbTiO$_{3}$.

\begin{table}[tb]
\caption{\label{tab_ramansus}Raman susceptibilities
and mode polarities (10$^{-2}$ a. u.)
of the A$_1$ TO modes in LiNbO$_3$, BaTiO$_3$ and PbTiO$_3$.}
\begin{ruledtabular}
\begin{tabular}{lrrrrrrrrrrrr}
         & &\multicolumn{3}{c}{LiNbO$_3$}& &\multicolumn{3}{c}{BaTiO$_3$}&
&\multicolumn{3}{c}{PbTiO$_3$} \\
         & &\multicolumn{1}{c}{$p_{3}$} &
            \multicolumn{1}{c}{$\alpha_{11}$} &
            \multicolumn{1}{c}{$\alpha_{33}$} & &
            \multicolumn{1}{c}{$p_{3}$} &
            \multicolumn{1}{c}{$\alpha_{11}$} &
            \multicolumn{1}{c}{$\alpha_{33}$} & &
            \multicolumn{1}{c}{$p_{3}$} &
            \multicolumn{1}{c}{$\alpha_{11}$} &
            \multicolumn{1}{c}{$\alpha_{33}$} \\
\hline
TO1 & & 3.65 &-0.70 & -2.02 & & 1.22 & -0.16 & -0.13 & & 1.25 &-0.67
& -0.43  \\
TO2 & & 0.45 & 0.30 &  0.53 & & 3.25 & -1.18 & -2.73 & & 2.18 &-0.75
& -0.33  \\
TO3 & & 0.67 & 0.18 & -0.05 & & 1.74 & -1.26 & -2.55 & & 2.68 &-2.42
& -2.28  \\
TO4 & & 3.82 &-1.96 & -3.23 & &      &       &       & &      &
&        \\
\end{tabular}
\end{ruledtabular}
\end{table}

The microscopic origin of the lower A$_1$ TO2
mode Raman susceptibility
in PbTiO$_3$, compared to BaTiO$_3$, is explained
by the decomposition of $\alpha_{33}$ into contributions of the individual
atoms in the unit cell (see Table \ref{tab_decomp}).
In both perovskites, the major contributions to the Raman
susceptibility of the A$_1$ TO2 modes are $\alpha_{33}(Ti)$ and
$\alpha_{33}(O_1)$ \cite{Note-positions};
$\alpha_{33}$ is mostly due
to the atomic displacements of the atoms located on the
Ti--O chains oriented along the polar direction. First,
the derivatives of $\chi^{(1)}_{33}$ versus atomic displacement are
of opposite sign for Ti and O$_{1}$ atoms, and significantly larger
in BaTiO$_3$ than in PbTiO$_3$. Second, the opposing displacements of
Ti and O$_{1}$ atoms in the TO2 mode in BaTiO$_{3}$ produce
contributions that add to yield a giant $\alpha_{33}$. On the
other hand, the in-phase displacements of Ti and O$_{1}$ in
PbTiO$_{3}$ produce contributions that cancel out, giving a
small $\alpha_{33}$. 
This distinct behaviour goes beyond a simple mass effect.
Changing the mass of Pb to that of Ba in the dynamical matrix of
PbTiO$_3$ has no significant effect on the relative Ti--O
displacement.
Large atomic
displacements of opposite direction along the Ti--O chains
are therefore needed to generate a large $\alpha_{33}$
and potentially a large $r_{33}$.

\begin{table}[tb]
\caption{\label{tab_decomp}
Decomposition of the Raman susceptibility of the A$_1$ TO2
mode in BaTiO$_3$ and PbTiO$_3$ into contributions from the
individual atoms in the unit cell~\cite{Note-positions}.
}
\begin{ruledtabular}
\begin{tabular}{ccrrrcrrr}
           & & \multicolumn{3}{c}{BaTiO$_3$} & & 
\multicolumn{3}{c}{PbTiO$_3$} \\
$\kappa$ & &
      $\sqrt{\Omega} \frac{\partial \chi^{(1)}_{33}}{\partial 
\tau_{\kappa 3}}$ &
      $u(\kappa,3)$ &
      $\alpha_{33}(\kappa)$ & &
      $\sqrt{\Omega} \frac{\partial \chi^{(1)}_{33}}{\partial 
\tau_{\kappa 3}}$ &
      $u(\kappa,3)$ &
      $\alpha_{33}(\kappa)$ \\
       & & (a.u.)  & \multicolumn{2}{c}{(10$^{-2}$ a.u.)}
       & & (a.u.)  & \multicolumn{2}{c}{(10$^{-2}$ a.u.)}      \\
\hline
Ba/Pb         & & 0.45 &-0.014 &-0.01 & &-1.00 &-0.006 & 0.01 \\
Ti            & &-6.46 & 0.257 &-1.66 & &-2.64 & 0.216 &-0.57 \\
O$_1$         & & 5.15 &-0.167 &-0.86 & & 3.69 & 0.059 & 0.22 \\
O$_2$/O$_{3}$ & & 0.43 &-0.240 &-0.10 & &-0.02 &-0.316 & 0.01 \\
\hline
Tot           & &      &       &-2.73 & &      &       &-0.32 \\
\end{tabular}
\end{ruledtabular}
\end{table}


In summary, we presented a method to compute
the EO tensor from first principles. 
In LiNbO$_3$ and BaTiO$_3$, the large EO
response originates in the giant contribution of the
successor of the soft mode, which combines
low frequency, high polarity and high Raman susceptibility.
In comparison, the contribution of the similar mode
in tetragonal PbTiO$_{3}$ is rather weak due to its low
Raman susceptibility. In the perovskites, the Raman
susceptibility is principally determined by the atomic displacements
along the B--O chains in the polar direction. 
This suggests
that the search for new perovskite oxides with good EO properties
should focus on compounds with large relative B--O atomic displacements
along the chains.

We thank M. Fontana, P. Bourson and B. Champagne
for discussions and K. M. Rabe for 
reading the manuscript. MV and XG acknowledge
FNRS Belgium. The work was supported by the Volkswagen Stif\-tung 
(Nano-sized ferroelectric Hybrids, I/77 737), FNRS
(9.4539.00 and 2.4562.03), the R\'egion Wallonne (Nomade,
115012), the PAI 5.01 and EU Exciting network.


\begin{thebibliography}{}
\bibitem{note_rc}
It is usual to define an effective EO
coefficient $r_c = r_{33} - (n_o/n_e)^3 r_{13}$ where $n_o$ and
$n_e$ are the ordinary and extraordinary refractive indices.
The theoretical values of ($n_o$,$n_e$,$r_c^{\eta}$) are:
LiNbO$_3$ (2.37, 2.35, 17.0),
BaTiO$_3$ (2.55, 2.42, 11.9),
PbTiO$_3$ (2.70, 2.61, -4.1).
The experimental values are:
LiNbO$_3$ (2.28, 2.19, 21.2),
BaTiO$_3$ (2.41, 2.36, 29.7),
PbTiO$_3$ (2.67, 2.66,-8.0).
\end{thebibliography}
\end{document}